\documentclass[graybox]{svmult}

\usepackage{mathptmx}       
\usepackage{helvet}         
\usepackage{courier}        
\usepackage{type1cm}        
\usepackage{makeidx}         
\usepackage{graphicx}        
\usepackage{multicol}        
\usepackage[bottom]{footmisc}

\newcommand{\beq}{\begin{equation}}
\newcommand{\eeq}{\end{equation}}
\newcommand{\bqa}{\begin{eqnarray}}
\newcommand{\eqa}{\end{eqnarray}}
\newcommand{\nn}{\nonumber}

\newcommand{\smallfrac}[2]{\mbox{$\frac{#1}{#2}$}}

\newcommand{\half}{\smallfrac{1}{2}}

\newcommand{\sq}[1]{\left[ {#1} \right]}

\newcommand{\tr}[1]{{\rm Tr}\sq{ {#1} }}

\makeindex             

\begin{document}

\title*{Trust-free verification of steering: why you can't cheat a quantum referee}
\author{Michael J. W. Hall}
\institute{M.J.W. Hall \at Centre for Quantum Dynamics, Griffith University, Brisbane, QLD 4111, Australia.\\ 
}
%
%
\maketitle

\abstract{It was believed until recently that the verification of quantum entanglement and quantum steering, between two parties, required trust in at least one of the parties and their devices, in contrast to the verification of Bell nonseparability. It has since been shown that this is not the case: the need for trust, in verifying two parties share a given quantum correlation resource, can be replaced by quantum refereeing, in which the referee sends quantum signals rather than classical signals to untrusted parties. The existence of such quantum-refereed games is discussed, with particular emphasis on how they make it impossible for the parties to cheat.  The example of a particular quantum-refereed steering game is used to show explicitly how measurement-device independence is achieved via `quantum programming' of untrusted  measurement devices;  how cheating is prevented by the steered party being unable  to distinguish sufficiently well between two sets of nonorthogonal signal states; and that cheating remains impossible when one-way communication is allowed from the steered party to the unsteered party.  This game has been recently implemented  experimentally, and is of particular interest both in accounting for any imperfections in the referee's preparation of signal states, and in suggesting the future possibility of secure two-sided quantum key distribution with Bell-local states.}

\section{Introduction}

Pure quantum states shared between two parties are rather simple with regard to possible types of correlations: they are either factorisable or entangled.  If they are factorisable, then all correlations are trivial.  If they are entangled then they are also Bell nonseparable, steerable, have quantum discord, etc.  Thus, the latter properties only become distinguishable for mixed quantum states. This was first clearly pointed out by Werner, who showed there are mixed quantum states that are both entangled and  Bell separable. \cite{werner}. Later it was shown that there are also mixed quantum states that both unentangled and discordant \cite{zurek}, and states that are both Bell separable and steerable \cite{wise2007}. In this way a hierarchical structure of quantum correlations has emerged, as reviewed in Sect.~\ref{hier}.

This hierarchy is of interest not just for foundational reasons: different types of quantum correlation reflect the resources needed to accomplish various tasks of physical interest.  Thus, for example, entanglement is necessary to optimally distinguish any two quantum channels \cite{chan}; steerability is necessary for subchannel discrimination \cite{sub} and allows one-sided secure key distribution \cite{1qkd}; and Bell nonseparability allows two-sided secure key distribution and randomness generation \cite{bellapp}.  It is therefore important to be able to verify or witness the level of correlation of a claimed resource.  This is typically done via testing for the violation of suitable inequalities, as is discussed in Sect.~\ref{wit}.

In Sect.~\ref{games}, it is recalled how verifying a given degree of correlation can be recast as a quantum correlation game, in which a referee sends signals to the parties, receives corresponding outputs from them, and calculates the value of a suitable payoff function from the correlations between the inputs and outputs.  Until recently, such games for verifying steering and entanglement were thought to require trust by the referee in at least one of the parties and their devices: a hierarchy of trust mirrored the hierarchy of correlations \cite{wisetrust}.  However, based on pioneering work by Buscemi, it is now known that trust can be replaced by `quantum refereeing', in which the referee sends quantum signals rather than classical signals to the parties \cite{busc,losr,branc,rosset,kocsis}.

Section~\ref{prevent} explores in depth how cheating by the parties is prevented in quantum-refereed games, using a steering game as an example. A formal proof is given for why the parties can only win this game if they share a steerable resource, before considering the physical reasons behind this.  It is shown explicitly how the impossibility of discriminating between sets of nonorthogonal signals from the referee prevents the success of possible cheating strategies.  It is also shown that the quantum signals from  the referee effectively `program' the measurement device of the receiving party, where the corresponding programs cannot be distinguished from one another, preventing any `hacking' by the parties.  Finally, it is shown that cheating remains impossible when one-way communication from the steered party to the steering party is permitted during the game.

Section~\ref{exp} considers experimental implementations of quantum-refereed correlation games, including the need for modification of payoffs due to imperfect preparation of signal states by the referee, and the robustness of these games when the signal states are transmitted through a noisy channel. Results from a recent experiment for a quantum-refereed steering game are briefly recalled \cite{kocsis}, that demonstrates trust-free verification of steering is in principle possible using a Bell-local resource.

Finally, a brief discussion is given in Sect.~\ref{con}.

\section{Hierarchy of quantum correlations}
\label{hier}

We consider the hierarchical structure of quantum correlations in more detail in this section, within a standard two-party scenario.  In particular, we assume that two distant parties, Alice and Bob,  can generate a set of statistical correlations in the following way.  On each run,  Alice makes some measurement labelled by $x$,  and obtains a result labelled by $a$.  Similarly, on each run Bob makes some measurement labelled by $y$, and obtains a result labelled by $b$.  Over many runs, therefore, they are able to estimate the set of joint probabilities $\{p(a,b|x,y)\}$. An aim of physics is to explain these joint probabilities and the statistical correlations that they generate. We will consider three types of physical explanation in particular.

\subsection{Entanglement and separability}

First, we can search for a separable quantum model of the correlations, where Alice's and Bob's measurement statistics are generated by a set of local quantum states on two Hilbert spaces $H_A$ and $H_B$ respectively. In particular, we say there is a separable quantum state model of the correlations on $H_A\otimes H_B$ if and only if they have the form
\beq \label{sep}
p(a,b|x,y) = \sum_\lambda\,p(\lambda)\,p_Q(a|x,\rho^A_\lambda)\,p_Q(b|y,\rho^B_\lambda) .
\eeq
Here $\lambda$ denotes a classical random variable with probability density $p(\lambda)$, $\rho^{A}_\lambda$ and $\rho^{B}_\lambda$ denote density operators on $H_{A}$ and $H_B$, and $p_Q(m|M,\rho)$ denotes a quantum probability distribution for state $\rho$ and measurement $M$, i.e,
\beq \label{pq}
p_Q(m|M,\rho) = \tr{E^M_x\rho}
\eeq
for some positive operator valued measure (POVM) $\{E^M_m\}$ (thus, $E^M_m\geq0$ and $\sum_m E^M_m=\hat 1$).

It follows that correlations with a separable quantum state model on $H_A\otimes H_B$ are equivalently described by the separable quantum state
\beq
\rho^{AB} := \sum_\lambda p(\lambda)\,\rho^A_\lambda\otimes\rho^B_\lambda
\eeq
on $H_A\otimes H_B$.  Conversely, correlations with no such separable quantum state model are defined to be {\it entangled} with respect to $H_A\otimes H_B$. 

\subsection{Steering and spooky action at a distance}

The concepts of entanglement and steering were introduced by Schr\"odinger~\cite{schr}, in his response to the famous Einstein-Podolsky-Rosen (EPR) paper of 1935~\cite{epr}.  In particular, he used `steering' to denote the property that, for a shared quantum state, Alice can typically control, via her choice of measurement, the corresponding set of local quantum states that Bob's system is described by. This steering of Bob's local state by a remote measurement is the `spooky action at a distance' that Einstein so disliked about quantum mechanics \cite{spooky}.

Clearly, there is no steering of the above type in the case that the statistical correlations between Alice and Bob can be explained via some fixed set of local quantum states for Bob: in this case Alice's measurements have no effect. A simple example is a factorisable state, $\rho^A\otimes\rho^B$, where Bob's local state is always described by $\rho^B$ independently of Alice's actions.  This consideration led
Wiseman {\it et al.} to formally define EPR-steering in terms of the existence or otherwise of a local hidden state (LHS) model for one of the parties \cite{wise2007}. 

In particular,  a given set of joint probabilities $\{p(a,b|x,y)\}$ is defined to have a local hidden state model for Bob, on Hilbert space $H_B$, if and only if
\beq \label{steer}
p(a,b|x,y) = \sum_\lambda\,p(\lambda)\,p(a|x,\lambda)\,p_Q(b|y,\rho^B_\lambda) .
\eeq
Here $p_Q(m|M,\rho)$ is a quantum probability distribution as per Eq.~(\ref{pq}), and $p(a|x,\lambda)$ can be an arbitrary probability distribution. Thus, all correlations are explained via some pre-existing set of local quantum states for Bob on $H_B$. Conversely, correlations that do not admit such an LHS model are defined to be {\it EPR steerable} from Alice to Bob, with respect to $H_B$~\cite{wise2007}. 

EPR steerability from Bob to Alice is similarly defined with respect to an LHS model for Alice, relative to some Hilbert space $H_A$. Thus, the concept of steering is inherently asymmetric.

Comparison of Eqs.~(\ref{sep}) and (\ref{steer}) show that, unlike separable state models of correlations, LHS models  do not require that the steering party (Alice in this case), is described by the laws of quantum mechanics.  They only require that there is some local statistical description for her outcomes, $p(a|x,\lambda)$.  Thus, for example, in any such model Bob's local statistics are subject to the Heisenberg uncertainty principle, but Alice's need not be.  This underlies tests for the existence of such models via steering inequalities \cite{cav2009}, as will be seen below.

\subsection{Bell nonseparability and local hidden variables}

The Einstein-Podolsky-Rosen paper of 1935 further inspired the consideration of an even more general class of correlation models: local hidden variable (LHV) models. In particular, a given set of joint probabilities $\{p(a,b|x,y)\}$ is defined to have a local hidden variable model if and only if 
\beq \label{lhv}
p(a,b|x,y) = \sum_\lambda\,p(\lambda)\,p(a|x,\lambda)\,p(b|y,\lambda) ,
\eeq
where both $p(a|x,\lambda)$ and $p(b|y,\lambda)$ can be arbitrary probability distributions. Conversely, if there is no such model, the correlations are said to be Bell nonseparable (or Bell nonlocal).

Such models were introduced by Bell~\cite{bell}, who famously showed how the nonexistence of such models for a given set of correlations can be tested experimentally, via what are now called Bell inequalities. Note that LHV models do not make any assumptions about how the local statistics are generated---in particular, unlike quantum separability and local quantum state models, there is no assumption that any particular theory, such as quantum mechanics, is valid.

\section{Witnessing the hierarchy}
\label{wit}

It is a logical consequence of the above definitions that joint quantum states on a given Hilbert space $H_A\otimes H_B$ have the hierarchical ordering
\beq
{\rm Bell~nonseparability~~\Longrightarrow~~EPR~steering~~\Longrightarrow~~entanglement},
\eeq
according  to the type of correlations they can generate via suitable measurements. This hierarchy is strict: there are steerable states that are not Bell nonseparable, and entangled states that are not steerable \cite{wise2007}.  A nice example is provided by the Werner states of two qubits, defined by 
\beq  \label{werner}
\rho_W := W\, |\Psi_-\rangle\langle\Psi_-| + (1-W)\,\frac{1}{4} \hat 1\otimes\hat 1 = \frac{1}{4}\left(\hat 1\otimes \hat 1- W\sum_j \sigma_j\otimes\sigma_j \right) ,
\eeq
where $|\Psi_-\rangle$ denotes the singlet state, $\sigma_1,\sigma_2,\sigma_3$ are the Pauli spin operators, and $-1/3\leq W\leq1$.  Werner states are Bell nonseparable for $W>1/\sqrt{2}$, EPR steerable for $W>1/2$, and entangled for $W>1/3$ \cite{werner,wise2007}.

Membership of each class in the hierarchy can be witnessed via suitable correlation inequalities. To see this, we consider the simplest case where Alice and Bob can each make two possible measurements, $x_1$ and $x_2$ for Alice and $y_1$ and $y_2$ for Bob, with each measurement having two possible outcomes $\pm1$.  We will denote the respective outcomes by $a_1,a_2,b_1$ and $b_2$.

First, if an LHV model as per Eq.~(\ref{lhv}) can predict the outcomes of each possible measurement (i.e., it is deterministic), it is easy to check that that these predetermined outcomes must satisfy
\[ a_1b_1 + a_1b_2 + a_2b_1 -a_2b_2 = a_1(b_1+b_2) + a_2(b_1-b_2) = \pm 2\]
for each run. Hence, the correlations satisfy the Bell inequality \cite{chsh}
\beq \label{chsh}
|\langle a_1b_1 \rangle +\langle a_1b_2 \rangle+\langle a_2b_1\rangle - \langle a_2b_2 \rangle | \leq 2.
\eeq
This inequality may similarly be shown to hold for nondeterministic LHV models of the correlations~\cite{chsh}, and is well known to be violated by some two-qbuit states (in particular, it is violated by two-qubit Werner states with $W> 1/\sqrt{2}$).

Second, if an LHS model for Bob on a qubit space, as per Eq.~(\ref{steer}), can predict the outcomes of Alice's possible measurements, and Bob's measurements $y_1, y_2$ correspond to measurements of $\sigma_1$ and $\sigma_2$ on his qubit, then for each local state $\rho^B_\lambda$ one has
\[
\langle a_1b_1\rangle_\lambda + \langle a_2b_2\rangle_\lambda =\tr{\rho^B_\lambda(a_1\sigma_1+a_2\sigma_2)} .\]
Now, the eigenvalues of the qubit operator $\pm\sigma_1+\pm\sigma_2$ are $\pm\sqrt{2}$ for any choice of the signs. Hence, averaging over $\lambda$, one obtains the EPR steering inequality \cite{cav2009}
\beq \label{steerin}
|\langle a_1\sigma_1\rangle + \langle a_2\sigma_2\rangle| \leq \sqrt{2}.
\eeq
This inequality easily generalises to the case that Alice's outcomes are not predetermined \cite{cav2009}. A simple calculation shows that it is violated, for example, by two-qbuit Werner states with $W>1/\sqrt{2}$ (a stronger steering inequality, violated for $W>1/\sqrt{3}$, will be given further below).

Third and finally, a quantum separable model for the correlations on a two-qbuit space, in the case that both Alice and Bob measure $\sigma_1$ and $\sigma_2$ (i.e., $x_j=y_j=\sigma_j$), implies via Eq.~(\ref{sep}) that
\[
\langle a_1b_1\rangle_\lambda + \langle a_2b_2\rangle_\lambda = \sum_{j=1}^2\tr{\rho^A_\lambda \sigma_j }\,\tr{\rho^B_\lambda \sigma_j } = \sum_{j=1}^2 m_j(\lambda)\,n_j(\lambda) ,
\]
where $m(\lambda)$ and $n(\lambda)$ denote the Bloch vectors of $\rho^A(\lambda)$ and $\rho^B(\lambda)$ respectively. Hence, since these Bloch vectors are at most of unit length, one obtains the entanglement witness inequality \cite{terhal}
\beq \label{entin}
|\langle \sigma_1\otimes\sigma_1\rangle + \langle \sigma_2\otimes\sigma_2\rangle| \leq \sum_\lambda p(\lambda)|m(\lambda)|\,|n(\lambda)|\leq 1,
\eeq
where the triangle and Schwarz inequalities have been used.  This inequality is violated by, for example, two-qubit Werner states with $W>1/2$.

\section{Quantum correlation games}
\label{games} 

It is possible to recast Bell, steering and entanglement inequalities, such as those in Eqs.~(\ref{chsh})-(\ref{entin}), into the form of games, played by Alice and Bob to convince a referee that they share a resource that is entangled, steerable, or Bell nonseparable. Alice and Bob are not allowed to communicate with each other during the game, although they can agree on a prearranged strategy beforehand.  On each run the referee, Charlie say, sends a measurement label $x$ to Alice, and receives a corresponding measurement outcome $a$ from Alice.  Similarly, Charlie sends a measurement label $y$ to Bob, and receives a corresponding measurement outcome $b$.  From many runs, Charlie can estimate the probabilities in the set $\{p(a,b|x,y)\}$, and determine whether they violate the inequality being tested.  

This inequality may also be used to calculate a suitable payoff, $\wp(a,b,x,y)$, to Alice and Bob on each run of the game. For example, consider the entanglement inequality in Eq.~(\ref{entin}).  In this case, a suitable payoff function is $\wp(a,b,x,y)=ab\,\delta_{xy}/p(x,y)-1$, where $p(x,y)$ denotes the joint probability that the referee sends $x$ to Alice and $y$ to Bob in any run.  The corresponding average payoff is therefore
\beq \label{payoff}
\bar\wp := \sum_{a,b,x,y}\wp(a,b,x,y)\,p(a,b|x,y)\,p(x,y) = \langle \sigma_1\otimes\sigma_1\rangle + \langle \sigma_2\otimes\sigma_2\rangle - 1 .
\eeq
Thus, Alice and Bob can only win the game, i.e., score a positive average payoff, if they can violate the entanglement inequality in Eq.~(\ref{entin}).

\subsection{Cheating in classically-refereed games: a hierarchy of trust}

For the case of Bell games, corresponding to the referee testing whether Alice and Bob share a Bell nonseparable resource, Charlie does not have to trust Alice or Bob, nor their measurement devices.  He may regard them as `black boxes', into which values of $x$ and $y$ are input, and from which values of $a$ and $b$ are output.  As long as the correlations between these inputs and outputs violate a Bell inequality, there is no LHV model that can explain them.  Bell games are said to be device independent.  

However, for steering games and entanglement games the situation is different.  Suppose, for example, that Alice and Bob claim that they share a two-qbuit entangled state that violates the entanglement inequality in Eq.~(\ref{entin}), whereas in fact they share no quantum state at all.  Can the referee be confident that they cannot win the corresponding correlation game?

The answer is no:  Alice and Bob (or their devices) can cheat. They can, for example, share a predetermined list of $+1$s and $-1$s, such as $\{1,-1, -1, 1, -1, \dots\}$, and on the $n$th run each return the $n$th member of the list as their output. In this way they will maximally violate the entanglement inequality in Eq.~(\ref{entin}), with a value of $2$ for the left hand side, and will obtain the maximal possible average payoff of $+1$ in Eq.~(\ref{payoff}).
The same cheating strategy will clearly also allow them to violate the steering inequality in Eq.~(\ref{steerin}).

For some time it was thought, therefore, that trust was required to verify entanglement and steering. In particular, entanglement inequalities are with respect to particular POVMs for each of Alice and Bob, such as in Eq.~(\ref{entin}) while steering inequalities are with respect to  particular POVMs for the steered party, such as in Eq.~(\ref{steerin}).  However, the referee has no mechanism for ensuring that these POVMs are actually measured to generate the reported outcomes, and so must simply trust this is the case.

Thus, until recently, the standard picture  was that tests of entanglement require trust in both Alice and Bob and their devices; tests of EPR steering require trust in the steered party and their device; and tests of Bell nonseparability require no trust at all \cite{wisetrust}. This picture places limitations on applications of quantum correlations.  For example, it implies that EPR-steering can only be used for one-sided secure key distribution, due to the need to trust the steered party \cite{1qkd}.

\subsection{Preventing cheating: quantum-refereed games}

Surprisingly, however, it turns out that the above hierarchy of trust can be dispensed with!  The idea, first proposed by Buscemi for the case of entanglement \cite{busc}, and elaborated on and generalised to steering by Cavalcanti {\it et al.} \cite{losr}, is for the referee to replace the need for trust by quantum channels.  

For example, in the case of verifying EPR steering, instead of sending a label $y$ to Bob via a classical communication channel, and  trusting him and his devices to implement the corresponding measurement, Charlie sends a  quantum state $\omega_y$ via a {\it quantum} channel. By choosing a suitable set of such states and a corresponding payoff function, this makes it impossible for Alice and Bob to demonstrate EPR steering unless they genuinely share a steerable state. Thus, they can outwit a classical referee, but not a quantum referee.

The underlying physical mechanism for overcoming cheating is that the quantum states sent by the referee are nonorthogonal.  Such states cannot be unambiguously distinguished, preventing Alice and Bob from knowing which correlation is being tested on a given run.  Together with a suitable payoff function, this completely undermines any cheating strategy for simulating quantum correlations that they do not actually share.  As will be seen in the next section, quantum refereeing may also be regarded as a means of `quantum programming' measurement devices: no one other than the referee knows precisely what instructions the devices have been given.

\section{Example: a quantum-refereed steering game}
\label{prevent}

Buscemi proved that a suitable quantum-refereed game exists for  verifying the entanglement of any given entangled state \cite{busc}.This was generalised by Cavalcanti {\it et al.} to prove the existence of a suitable quantum-refereed game for verifying the steerability of any given EPR-steerable state \cite{losr}.  However, these existence proofs gave no explicit method for constructing such games.  This was remedied for entanglement games by Branciard {\it et al.}, who showed how to construct a quantum-refereed game for each possible entanglement witness inequality \cite{branc}. Remarkably, Rosset {\it et al.}  showed that one can even construct quantum-refereed entanglement games in which the requirement of no communication between Alice and Bob can be removed: such communication does not enable them to cheat \cite{rosset}.

Kocsis {\it et al.} similarly showed how to construct a suitable quantum-refereed steering game that tests for violation of a given EPR steering inequality.  An example of such a game is discussed in this section, with an emphasis on understanding why the game requires no trust in either of Alice and Bob (even when Bob is permitted one-way communication with Alice).  

\subsection{Rules of the game}
\label{game}

We now consider the following example of a quantum-refereed steering game. On each run of the game, Charlie sends Alice a classical signal $j\in {1,2,3}$, and Bob a qubit signal corresponding to an eigenstate of $\sigma_j$, i.e., a density operator $\omega^C_{j,s}:=(1/2)(\hat 1+s \sigma_j)$ with $s=\pm 1$.  Alice is required to return a value $a=\pm1$, while Bob is required to return a value $b=0$ or $1$. The average payoff function is defined to be
\beq \label{pay}
\bar \wp := 2 \sum_{j,s} \left( s\langle a b\rangle_{j,s} -\frac{1}{\sqrt{3}}\langle b\rangle_{j,s}\right),
\eeq
where $\langle\cdot\rangle_{j,s}$ denotes the average over those runs with a given value of $j$ and $s$.
The game is won if Alice and Bob can achieve an average payoff $\bar\wp>0$.

As per the general proof of Kocsis {\it et al.}, and as shown directly for this particular game below, Alice and Bob can win only if they genuinely share a steerable state \cite{kocsis}.  In fact, as will be shown below, they can  win only if they can violate the known EPR steering inequality \cite{cav2009}
\beq \label{cav3}
\langle a_1\sigma_1\rangle + \langle a_2\sigma_2\rangle + \langle a_3\sigma_3\rangle \leq \sqrt{3},
\eeq
where $a_j$ denotes Alice's outcome for input signal $j$. Indeed, we will see that the average payoff function is equal to the amount of violation of this inequality that they can achieve with a two-qbuit shared state.  Note that this steering inequality is a simple generalisation of the one in Eq.~(\ref{steerin}), and is proved the same way, noting that the eigenvalues of the qubit operator $\pm\sigma_1\pm\sigma_2\pm\sigma_3$ are $\pm\sqrt{3}$ for any choice of signs.

\subsection{Why cheating is impossible}
\label{proof}

Suppose that Alice and Bob do not share a steerable resource.  Hence, by definition, there must be an LHS model for Bob on some Hilbert space $H_B$ (not necessarily a qubit space), i.e., all correlations between Alice and Bob are described by a model as per Eq.~(\ref{steer}) for some ensemble of hidden states $\rho^B_\lambda$ on some Hilbert space $H_B$.  Now, since Bob receives an unknown state from Charlie, the most general action he can take to return a value $b=0$ or 1 is to measure some POVM $\{E^{BC}_0,E^{BC}_1\}$ on the combination of his local hidden state and the received state, and return the outcome. Note this includes, for example, strategies such as first making a measurement on the unknown state to try and determine the value of $j$ and $s$ and then making a corresponding measurement on his local state. 

It follows that the average value of $ab$, when Charlie sends Alice $j$ and Bob the state $\omega^C_{js}$, is given by
\begin{eqnarray}
\langle ab\rangle_{j,s} &=& \sum_\lambda p(\lambda) \,\langle a\rangle_{j,\lambda}\, \langle b\rangle_{j,s,\lambda} \nn \\
&=& \sum_\lambda p(\lambda) \,\langle a\rangle_{j,\lambda}\, {\rm Tr}_{BC}[E^{BC}_1\rho^B_\lambda\otimes\omega^C_{js}]\nn \\
&=& \sum_\lambda p(\lambda) \,\langle a_j\rangle_{\lambda}\, {\rm Tr}[X^{C}_\lambda\omega^C_{js}], \nn
\end{eqnarray}
where we rewrite $\langle a\rangle_{j,\lambda}$ as $\langle a_j\rangle_{j,\lambda}$ (i.e., $a_j=\pm 1$ denotes Alice's outcome for input~$j$), and define the positive operator $X^C_\lambda$ on $H_C$ by $X^C_\lambda:= {\rm Tr}_{B}[E^{BC}_1\rho^B_\lambda\otimes\hat 1_C]$.  Defining the density operator $\tau^C_\lambda$, probability density $q(\lambda)$ and positive constant $N$ by
\[ \tau^C_\lambda:= X^C_\lambda/\tr{X^C_\lambda},\qquad q(\lambda):= p(\lambda) \tr{X^C_\lambda}/N,\qquad N:=\sum_\lambda  p(\lambda) \tr{X^C_\lambda}, \]
then yields 
\beq 
\langle ab\rangle_{j,s} = N \sum_\lambda q(\lambda)\, \langle a_j\rangle_\lambda\, \tr{\omega^C_{js}\tau^C_\lambda}.
\eeq
One similarly finds
\beq
\langle b\rangle_{j,s} = N \sum_\lambda q(\lambda) \,\tr{\omega^C_{js}\tau^C_\lambda}.
\eeq
Hence, noting from the definition of the states $\omega^C_{js}$ in Sect.~\ref{game} that $\sum_s s\,\omega^C_{js}=\sigma_j$ and $\sum_s \omega^C_{js}=\hat 1$, the value of the average payoff in Eq.~(\ref{pay}) is given by
\begin{eqnarray}
\bar\wp &=& 2N \sum_{\lambda,j} q(\lambda) \left( \langle a_j\rangle_\lambda\,\tr{\sigma_j\tau^C_\lambda} - \frac{1}{\sqrt{3}}\right) \nn \\
&=& 2N  \left[ \sum_j \langle a_j\sigma_j\rangle_{\rm LHS} - \sqrt{3} \right],
\end{eqnarray}
where the average is with respect to the LHS model defined by the probability density $q(\lambda)$ and the corresponding hidden states $\tau^C_\lambda$ on Charlie's qubit Hilbert space.  Hence, using the steering inequality in Eq.~(\ref{cav3}), $\bar \wp\leq 0$, i.e., Alice and Bob cannot with the game with a nonsteerable resource, as claimed.

The above proof that Alice and Bob cannot cheat is, necessarily, somewhat formal in nature.  The focus in the remainder of this section is on giving some physical insight into why cheating is impossible, and also showing how Alice and Bob can to win the game if they do share a suitable steering resource.

\subsection{Connection to unambiguous state discrimination}
\label{unambig}

Some insight is gained by considering a possible cheating strategy that Alice and Bob could employ if they share no quantum state.  It is clear from Eq.~(\ref{pay}) that the average payoff is maximised if Bob returns the outcome $b=0$ whenever $s a=-1$.  Now, Alice has no access to the value of $s$ (she is only sent the value of $j$), but Bob can in principle try to estimate $s$ from the state $\omega^C_{js}$ sent to him by the referee.  Hence, an obvious cheating strategy is for Alice to always return the result $a=1$, and for Bob to return the value $b=1$ if his estimated value of $s$ is 1 and $b=0$ otherwise.  
Note that this strategy can also be easily varied to the case where Alice returns $a=\pm1$ according to a preagreed list, while Bob returns $b=1$ if and only if his estimated value of $s$ equals this value. This variation allows Alice to return seemingly random outputs, while yielding the same average payoff.

If Bob can precisely determine the value of $s$, the above strategy results in the maximum possible average payoff,
\begin{equation} \label{precise}
\bar\wp = 2 \sum_j \left( 1 - \frac{1}{\sqrt{3}}\right) = 2(3-\sqrt{3}).
\eeq
More generally, if $p(+|s,j)$ denotes the probability that Bob estimates $s=+1$ when the state $\omega^C_{js}$ is sent by the referee, then the strategy yields 
\[
\sum_s s\langle b\rangle_{j,s} = p(+|+,j) - p(+|-,j),\qquad \sum_s \langle b\rangle_{j,s} = p(+|+,j) + p(+|-,j), \]
and hence  the average payoff is given by
\begin{eqnarray}
\bar\wp &=& 2 \sum_j\left[ \left( 1 - \frac{1}{\sqrt{3}}\right) p(+|+,j) - \left( 1 + \frac{1}{\sqrt{3}}\right) p(+|-,j) \right] \nn\\
&=& 6 \left[ \left( 1 - \frac{1}{\sqrt{3}}\right) \bar p(+|+) - \left( 1 + \frac{1}{\sqrt{3}}\right) \bar p(+|-) \right].
\end{eqnarray}
Here, $\bar p(+|+):=(1/3)\sum_j p(+|+,j)$ is the average probability that Bob correctly identifies $s=1$, while $\bar p(+|-):=(1/3)\sum_j p(+|-,j)$ is the average probability that Bob wrongly identifies $s=1$, i.e., a false positive.

It follows immediately that the condition for this cheating strategy to be successful is that the ratio of true positives to false positives satisfies
\beq \label{ch1}
\frac{\bar p(+|+)}{\bar p(+|-)} > \frac{ \sqrt{3}+1}{\sqrt{3}-1} \approx 3.732 .
\eeq
This might not appear too much to ask.  But, in fact, it is impossible.  Bob's aim is to successfully distinguish the set of states $\{\omega^C_{j+}\}$ from the set of states $\{\omega^C_{j-}\}$, where these two sets are clearly not mutually orthogonal.  Unfortunately for Bob, quantum mechanics places strong constraints on the success with which such sets can be unambiguously distinguished. 

In particular, for Bob to estimate the value of $s$, he will have to measure some POVM $\{M_\pm\}$ on the states sent to him by Charlie.  It follows, recalling that $\sigma^C_{js}=\half(1+s\sigma_j)$, that 
\[ \bar p(+|s) =  (1/3)\sum_j \tr{M_+\omega^C_{js}} = (1/6) \sum_j \tr{M_+(1+s\sigma_j)}. \]
Further, the requirement $M_+>0$ implies that $M_\pm=\mu (1+m\cdot \sigma)$ for some $\mu>0$ and 3-vector $m$ of length no greater than unity.
Substitution then gives
\beq \label{ch2}
\frac{\bar p(+|+)}{\bar p(+|-)} = \frac{3 +\sum_j m_j}{3-\sum_j m_j} \leq \frac{ \sqrt{3}+1}{\sqrt{3}-1} , 
\eeq
where the inequality is easily obtained by maximising $\sum_j m_j$ subject to the constraint $m\cdot m\leq 1$ (equality corresponds to $m_j\equiv 1/\sqrt{3}$).

Comparison of Eqs.~(\ref{ch1}) and (\ref{ch2}) immediately shows that the cheating strategy fails:  Bob cannot make any measurement that distinguishes the input states sufficiently for him to estimate $s$ to the required degree of accuracy.

\subsection{Quantum refereeing as quantum programming of measurement devices}

It also of interest to give some insight as to how Alice and Bob can win this quantum-refereed steering game, when they do share a suitable steerable quantum state $\rho$.  In particular, the states sent by the referee can be regarded as `programming' Bob's devices to make corresponding measurements, where neither Bob nor his devices are able to cheat by `reading' the program (as this would again correspond to distinguishing between sets of nonorthogonal states).

In the general case, suppose that Alice measures the POVM $\{E^x_a$\} on receipt of classical input $x$ from the referee, and Bob measures the POVM $\{E^{BC}_b\}$ on receipt of quantum input $\omega^C_y$ from the referee.  The joint outcome probability distribution corresponding to $x$ and $y$ then follows as
\beq
p(a,b|x,y) = \tr{(E^x_a\otimes E^{BC}_b) (\rho\otimes \omega^C_{y})} = \tr{ (E^x_a\otimes M^{y}_{b}) \rho},
\eeq
where $M^{y}:=\{ M^{y}_b\}$ is an `induced'  or `programmed' POVM on Bob's Hilbert space, defined by
\beq \label{m}
M^{y}_b := {\rm Tr}_C[E^{BC}_b \omega^C_{y}].
\eeq
Thus, unlike a classically-refereed game, in which the referee sends a classical signal $y$ and Bob chooses a corresponding measurement to make on his system, a quantum referee sends a quantum signal $\omega^C_{y}$ that determines Bob's corresponding measurement on his shared system: Bob's measurement is `quantum programmed' by the referee (although Bob retains some freedom via his choice of $\{E^{BC}_b\}$).

For the particular quantum-refereed steering game defined in Sect.~\ref{game}, consider the case where $E^{BC}_1$ corresponds to the projection onto a singlet state, i.e., 
\beq \label{ebc}
E^{BC}_1 = |\psi_-\rangle\langle\psi_-| =  \frac{1}{4}\left(\hat 1\otimes \hat 1- \sum_j \sigma_j\otimes\sigma_j \right).
\eeq
Thus, Bob makes a partial Bell-state measurement on the combination of the state sent by Charlie and his component of the state $\rho$ that he shares with Alice.  Note that Bell-state measurements are natural in the context of quantum-refereed correlation games, as they are an integral part of the existence proofs by Buscemi~\cite{busc} and Cavalcanti {\it et al.}~\cite{losr}. The adequacy of {\it partial} Bell-state measurements was discovered by Branciard {\it et al.} for entanglement games \cite{branc}, and generalised to steering games by Kocsis {\it et al.}~\cite{kocsis}.  It can be shown more generally that it is always best for Bob to make an entangling (i.e., non-factorisable) measurement.

Equations (\ref{m}) and (\ref{ebc}) yield the corresponding programmed POVM element
\beq
M^{js}_1 = \frac{1}{8} {\rm Tr}_C\left[\left(\hat 1\otimes \hat 1- \sum_k \sigma_k\otimes\sigma_k\right)\,(\hat1+s\sigma_j)\right] = \frac{1}{4}(\hat1-s\sigma_j) = \half \omega^C_{j,-s} .
\eeq
Thus, when an eigenstate of $\sigma_j$ is sent by the referee, the programmed POVM element is proportional to the projection onto the orthgonal eigenstate.  In this way the referee effectively receives information about measurements of $\sigma_j$ on Bob's component of the shared state. Bob cannot cheat, however, because he does not know  which $\sigma_j$ measurement is actually programmed in any run.

If Alice further measures $-\sigma_j$ on receipt of signal $j$ from the referee, and they share a Werner state as in Eq.~(\ref{werner}), then substitution into Eq.~(\ref{pay}) yields the average payoff \cite{kocsis}
\beq
\bar\wp = -\sum_j \tr{\sigma_j\otimes\sigma_j\rho_W} - \sqrt{3} = 3W - \sqrt{3}. 
\eeq
Thus, Alice and Bob can win the game for any value $W>1/\sqrt{3}$.  

More generally, it is not difficult to show that whatever measurement Alice makes, the average payoff for the game under a partial Bell-state measurement by Bob corresponds precisely to the degree by which the corresponding steering inequality in Eq.~(\ref{cav3}) is violated.  

\subsection{Relaxing communication restrictions?}
\label{relax}

As mentioned previously, Rosset {\it et al.} have demonstrated the existence of quantum-refereed entanglement games for which the requirement that Alice and Bob do not communicate during the game can be relaxed \cite{rosset}.  Here the extent to which a similar relaxation is possible for quantum-refereed steering games is investigated, for the example in Sect.~\ref{game}.  It turns that that while allowing communication from Alice to Bob permits cheating, allowing communication from Bob to Alice does not.

In particular,  for the steering game in Sect.~\ref{game}, suppose first that Alice can send classical signals to Bob.  They can then cheat as follows:  Alice passes the input $j$ she receives from the referee on to Bob.  Bob uses this information to measure $\sigma_j$ on the corresponding state $\omega^C_{js}$ he has received from the referee, thus determining $s$. This in turn allows a perfect implementation of the cheating strategy in Sect.~\ref{unambig}, yielding a positive average payoff of $2(3-\sqrt{3})$ as per Eq.~(\ref{precise}).  Hence, one-way communication from Alice to Bob cannot be permitted in this game.

Conversely, however, suppose that Bob can send classical signals to Alice, but not vice versa.   From the form of the average payoff in Eq.~(\ref{pay}), the only way to cheat is for Alice to ensure that her output satisfies $a=sb$ as often as possible, so that a positive sign dominates in the first term.  Since Bob can send her his value of $b$, she therefore only further needs a reliable estimate of $s$ from him.  But, as we have already seen in Sect.~\ref{unambig}, Bob's best possible estimate of $s$ is just not good enough to be of any help to her.  Hence, one-way communication from Bob to Alice, i.e., from the steered party to the steering party, can be permitted without compromising the game.

\section{Experiment: trust-free verification of EPR steering}
\label{exp}

A number of experimental implementations of quantum-refereed entanglement games have now been performed \cite{ent1,ent2,ent3}, as well as an  experimental implementation of a quantum-refereed steering game \cite{kocsis}.  The latter is of particular interest for two reasons: (i) it provides a proof of principle for trust-free verification of EPR steering, raising the possibility of two-sided secure quantum key distribution without the need for a Bell nonseparable state; and (ii) it explicitly accounts for imperfections in the referee's preparation of the states he sends to Bob.  

In particular, in any experimental implementation of a quantum-refereed game, the referee cannot perfectly prepare the states intended to be sent to the untrusted party or parties.  The accuracy of preparation must therefore be taken into account to prevent cheating by Alice and Bob.  As an extreme example, suppose for the steering game in Sect.~\ref{game} that the referee prepares the states $\tilde\omega^C_{js}:=\half(1+s\sigma_1)$, independently of $j$, rather than eigenstates of $\sigma_j$.  Then Bob can unambiguously determine $s$ by measuring $\sigma_1$, and then implement the cheating strategy of Sect.~\ref{unambig} to achieve a positive average payoff of $2(3-\sqrt{3})$.

The way to account for imperfect preparations is to modify the payoff function for the game, based on tomography of the prepared states.  For example, in Ref.~\cite{kocsis} the modified average payoff
\beq \label{modpay}
\bar\wp(r):=  2 \sum_{j,s} \left( s\langle a b\rangle_{j,s} -\frac{r}{\sqrt{3}}\langle b\rangle_{j,s})\right)
\eeq
was used, where $r\geq 1$ is a tomographically-determined measure of the imperfect preparation, with $r=1$ corresponding to perfect preparation.  Recalling $b\geq 0$, one has $\bar\wp(r)<\bar\wp(1)$ for $r>1$, and so a positive average payoff is harder to achieve for imperfect state preparation---indeed so much harder that Alice and Bob are prevented form cheating, as follows from an argument analogous to that in Sect.~\ref{proof}~\cite{kocsis}.

It should further be noted that any noise, in the quantum channel used to send states to the untrusted parties, does not compromise quantum-refereed games \cite{kocsis}.  For example, for the steering game in Sect.~\ref{game}, suppose that Bob receives the states $\phi(\omega^C_{js})$, where $\phi$ is a completely positive trace preserving (CPTP) map describing the quantum channel used by the referee.  To show this does not allow any cheating by Alice and Bob, note first that
\beq  {\rm Tr}_{BC}[E^{BC}_b\rho^B_\lambda\otimes\phi(\omega^C_{js})] =  {\rm Tr}_{BC}[\tilde E^{BC}_b\rho^B_\lambda\otimes\omega^C_{js}] ,
\eeq
for any joint POVM $\{E^{BC}_b\}$ on $H_B\otimes H_C$, where the modified POVM $\{\tilde E^{BC}_b\}$ is defined by $\tilde E^{BC}_b:=(I_B\otimes \phi^*)(E^{BC}_b)$.  Here $I_B$ denotes the identity map on $H_B$, and $\phi^*$ denotes the dual map defined by $\tr{X\phi(Y)}=\tr{\phi^*(X)Y}$.  It is easily checked from this definition that $\{\tilde E^{BC}_b\}$ is indeed a POVM.  Hence, 
the proof in Sect.~\ref{proof} that Alice and Bob cannot cheat goes through just as before, using the modified POVM $\{\tilde E^{BC}_b\}$ in place of $\{E^{BC}_b\}$.

It is worth noting that this robustness of quantum-refereed games under noise may reduce the degree to which the average payoff needs to be modified to account for imperfect preparation.  For example, if the experimentally prepared states $\tilde\omega^C_y$ can be written in terms of the intended states as $\tilde\omega^C_y=\phi(\omega^C_y)$, for some CPTP map $\phi$, then no modification at all is necessary.  More generally, however, some tradeoff between finding a suitable $\phi$ and modifying the payoff will be necessary.

The experiment reported by Kocsis {\it et al.} implemented the quantum-refereed steering game in Sect.~\ref{game} using optical polarisation qubits and partial Bell-state measurements, where imperfect state preparation by the referee required a modifed average payoff $\bar\wp(r)$ as per Eq.~(\ref{modpay}), with $r=1.081$ \cite{kocsis}. Positive average payoffs of $1.09\pm 0.03$ and $0.05\pm0.04$ were obtained, for shared Werner states with $W=0.98$ and $W=0.698$ respectively, thus confirming a shared steering resource without any trust in Alice and Bob.  The latter case is of particular interest, as this Werner state does not violate any known Bell inequality, including the Bell inequality in Eq.~(\ref{chsh}) (which requires $W\geq 1/\sqrt{2}\approx 0.707$).

\section{Discussion}
\label{con}

Quantum-refereed games remove the need for trust in parties and their devices, when verifying the correlation strength of a shared resource, by replacing trust with quantum signal states. The physical mechanism by which cheating is prevented, including when communication restrictions are relaxed, has been studied in detail in Sect.~\ref{prevent}, using a particular quantum-refereed steering game as an example.

It should be noted that while quantum refereeing regains measurement device independence, in the verification of any level of the hierarchy of quantum correlations, this is at some cost.  First, while there is no need to trust Alice and Bob or their devices, the referee must be able to trust his own characterisation of the quantum states he sends.  Second, for Alice and Bob to win a quantum-refereed game, the parties that are sent a quantum state by the referee must be able to perform a joint measurement on that state and their local state. Hence, the technical demands are higher than for tests of Bell nonseparability, in which only classical signals need be sent. Finally, as is seen in the proof given in Sect.~\ref{proof}, the referee must trust that Alice and Bob's devices are subject to the laws of quantum mechanics (although no particular quantum model of the devices need be assumed).

Future work includes exploring whether allowing one-way communication from the steered party to the steering party, as per the example in Sect.~\ref{relax},  can be generalised to all quantum-refereed steering games; and finding an explicit protocol for secure two-sided quantum key distribution based on a steerable but Bell-nonlocal resource.  It would also be of interest to investigate whether proposed measures of steerability of quantum states in the literature \cite{sub,pro1,pro2} respect the ordering induced by quantum-refeered steering games defined in Ref.~\cite{losr}.

\begin{acknowledgement}
This work was supported by the Australian Research Council, Project No. DP 140100648. I thank Cyril Branciard, Francesco Buscemi, Yeong-Cherng Liang, Geoff Pryde, Dylan Saunders, Ernest Tan and Howard Wiseman for useful discussions.
\end{acknowledgement}

\end{document}